\documentclass[sigconf]{acmart}
\usepackage{balance}

\def\BibTeX{{\rm B\kern-.05em{\sc i\kern-.025em b}\kern-.08emT\kern-.1667em\lower.7ex\hbox{E}\kern-.125emX}}

\copyrightyear{2020}
\acmYear{2020}
\setcopyright{rightsretained}
\acmConference[ICPE '20]{Proceedings of the 2020 ACM/SPEC International Conference on Performance Engineering}{April 20--24, 2020}{Edmonton, AB, Canada}
\acmDOI{10.1145/3358960.3375791}

\theoremstyle{plain}
\newtheorem{problem}{Problem}

\begin{document}
\newcommand{\todo}[1]{}

\title[Change Point Detection in Software Performance Testing]{The Use of Change Point Detection to Identify Software Performance Regressions
  in a Continuous Integration System}
\author{David Daly}
\orcid{0000-0001-9678-3721}
\affiliation{MongoDB Inc}
\email{david.daly@mongodb.com}

\author{William Brown}
\affiliation{Columbia University}
\email{w.brown@columbia.edu}

\author{Henrik Ingo}
\orcid{0000-0003-1571-5108}
\affiliation{MongoDB Inc}
\email{henrik.ingo@mongodb.com}

\author{Jim O'Leary}
\orcid{0000-0002-3923-5742}
\affiliation{MongoDB Inc}
\email{jim.oleary@mongodb.com}

\author{David Bradford}
\orcid{0000-0003-2282-6535}
\affiliation{MongoDB Inc}
\email{david.bradford@mongodb.com}

\begin{abstract}
  We describe our process for automatic detection of performance changes for a software product in
  the presence of noise. A large collection of tests run periodically as changes to our software
  product are committed to our source repository, and we would like to identify the commits
  responsible for performance regressions. Previously, we relied on manual inspection of time series
  graphs to identify significant changes. That was later replaced with a threshold-based detection
  system, but neither system was sufficient for finding changes in performance in a timely
  manner. This work describes our recent implementation of a change point detection system built
  upon the E-Divisive means~\cite{edivisive2014} algorithm. The algorithm produces a list of change
  points representing significant changes from a given history of performance results. A human
  reviews the list of change points for actionable changes, which are then triaged for further
  inspection. Using change point detection has had a dramatic impact on our ability to detect
  performance changes. Quantitatively, it has dramatically dropped our false positive rate for
  performance changes, while qualitatively it has made the entire performance evaluation process
  easier, more productive (ex. catching smaller regressions), and more timely.
\end{abstract}



\maketitle

\section{Introduction}
We work in a software and services company and need to understand the performance of the software we
develop and sell. Our continuous integration system runs thousands of benchmarks periodically
(most commonly every 2 hours or every 24 hours), which each produce one or more scalar values as a
result. It is a challenge to analyze all of those results and historical data to determine whether the test should
be considered passed or failed. It is inherent to this type of benchmarking that the results contain
some level of noise. In most of our tests the worst case run to run variation level is currently less than $10\%$, but some
sensitive tests can fluctuate as much as $20\%$ or more over the course of a small numbers of runs. In this paper, we detail the results of our
experience in deploying automated change point detection software for identifying performance
changes in an evolving code base.

\subsection{Existing performance results and analysis}\label{subsec:existing}

Our performance testing infrastructure is built upon our continuous integration (CI) system:
Evergreen~\cite{Erf2016, Evergreen}. Evergreen tracks every commit to our source repository, compiles
the software on a number of different build targets, and runs correctness tests.

Evergreen also runs our performance tests. The performance tests take longer to run than the
correctness tests, so we run them less frequently, but otherwise the performance and correctness tests are the same from the
perspective of Evergreen.

With any performance test, we must
\begin{itemize}
\item Setup a system under test
\item Run a workload against the system under test
\item Report the results from the test
\item Decide (and alert) if the performance changed
\item Visualize the results
\end{itemize}

\begin{figure}[ht]
  \centering
  \includegraphics[width=\columnwidth]{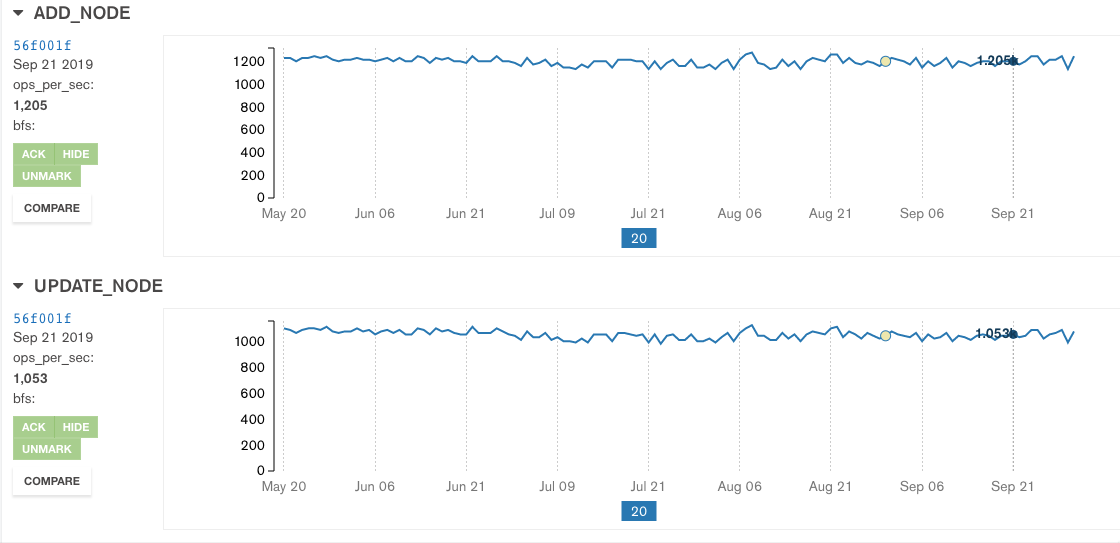}
  \caption{\label{fig:trend-linkbench} Performance trend graph with noise and no discernible changes.}
\end{figure}

\begin{figure}[ht]
  \centering
  \includegraphics[width=\columnwidth]{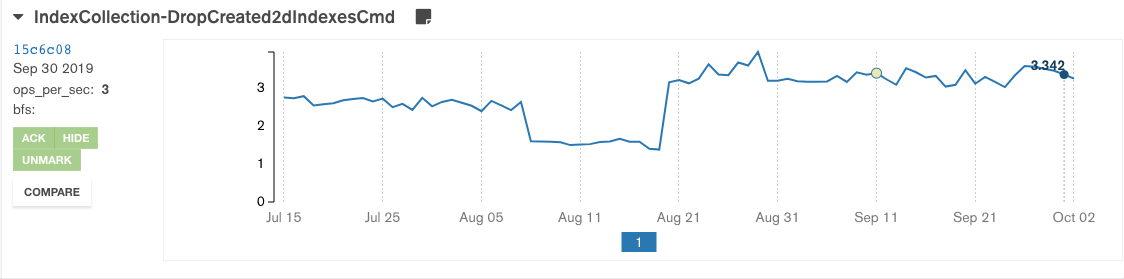}
  \caption{\label{fig:trend-index2} Example of a clear drop in performance on August 8th and subsequent
    performance recovery on August 20th.}
\end{figure}

\begin{figure}[ht]
\centering
\includegraphics[width=\columnwidth]{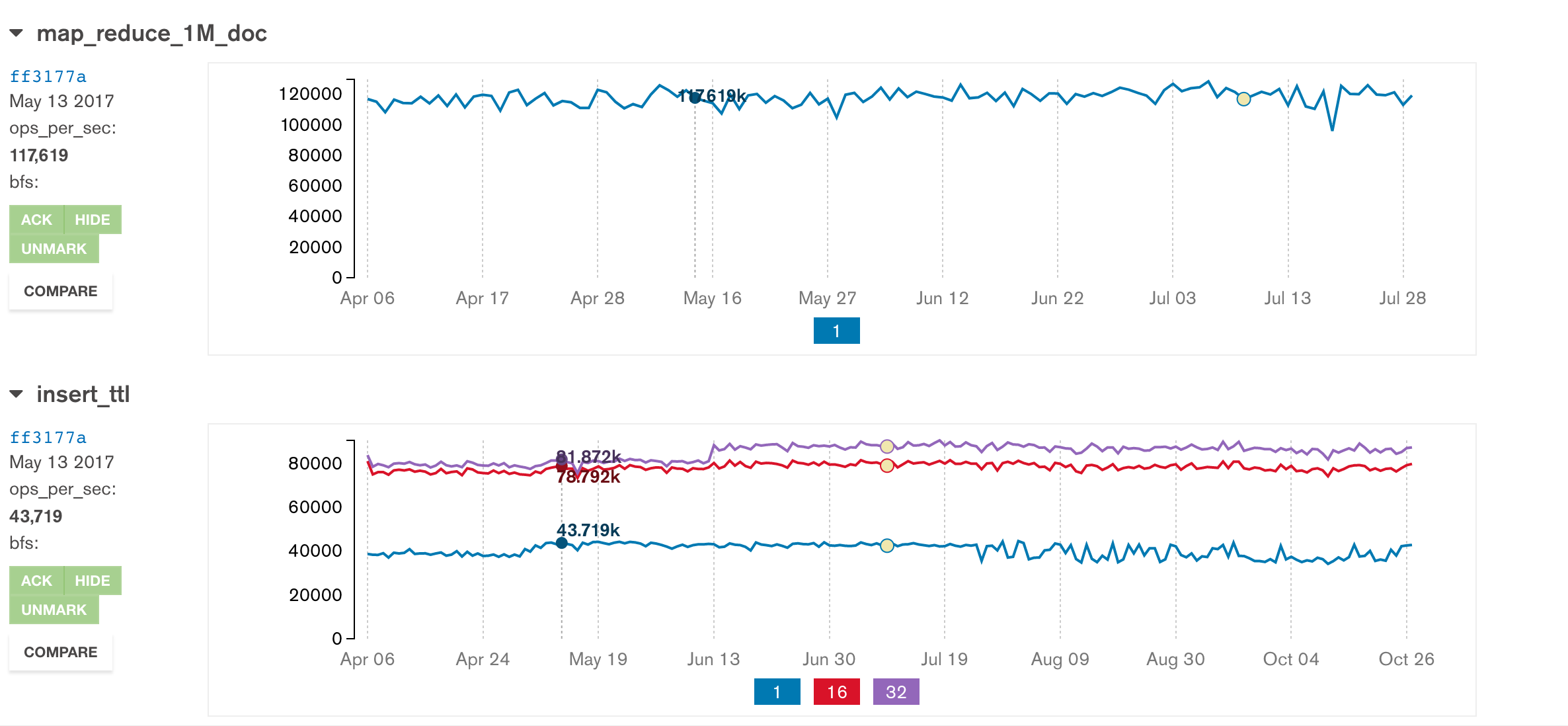}
\caption{\label{fig:trend-map} Performance trend graphs with noise. The second graph (insert\_ttl) has a change
  in performance on June 13th smaller than the noise threshold of the first graph. The amount of noise also
  changes over time for the insert\_ttl graph.}
\end{figure}

\begin{figure}[ht]
\centering
\includegraphics[width=\columnwidth]{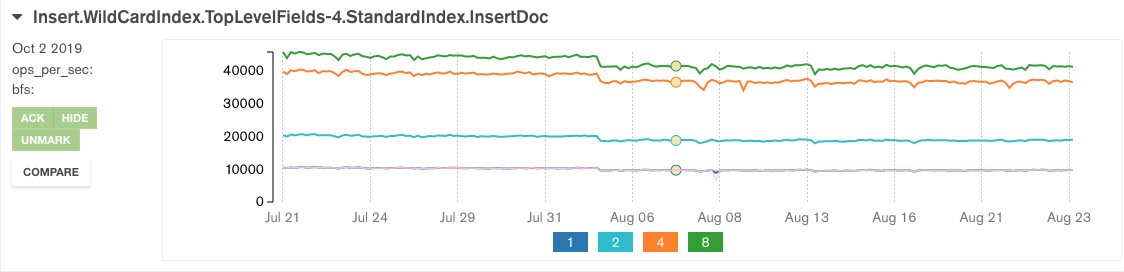}
\caption{\label{fig:wildcard-clean} Performance trend graph with a small drop in performance on
  August 5th. It
  would be easy for a person to miss that regression when looking through many graphs.}
\end{figure}

This paper focuses on the last two bullets\footnote{Previous work focused on the first two bullets
and getting reproducible results~\cite{Ingo2019}}. When we first setup our performance tests in our CI
system, we had a visualization of the performance over time, but no automated alerting. We had a
person (called the ``Build Baron'') looking through the graphs for regressions, and opening Jira
tickets\footnote{Atlassian's Jira is our ticketing system.} for any changes requiring further investigation. The Build Baron would be looking at performance trend graphs
like those shown in Figure~\ref{fig:trend-linkbench} and
Figure~\ref{fig:trend-index2}. Figure~\ref{fig:trend-linkbench} shows a test with some basic noise,
but no real changes in performance, while Figure~\ref{fig:trend-index2} has two very clear changes
in performance. The cases in
Figures~\ref{fig:trend-linkbench} and~\ref{fig:trend-index2} should be easy for the Build Baron to
triage, but many other cases are
not, such as the two graphs in Figure~\ref{fig:trend-map} and the graph in
Figure~\ref{fig:wildcard-clean}, which have smaller changes and larger run to run variation. We quickly realized that having humans triaging performance changes solely on the trend graphs was not a tenable solution: humans looking through all the graphs lose
focus quickly and the problem would get worse as we added
more tests and system configurations. Therefore we added an automated detection system.

The automated detection system was designed to catch
\begin{itemize}
\item Sudden drops in performance from a specific software change
\item Multiple small changes over time causing a drift down in performance
\end{itemize}
It did this in a very simple manner by doing three comparisons:
\begin{enumerate}
\item Compare to the immediately preceding run. Did it change more than $X\%$\footnote{Where a human
    then looked at flagged failures and $X$ is a parameter of the system. We used $10\%$ by default}?
\item Compare to a run from a week ago. Did it change more than $X\%$?
\item Compare to a run for the last stable release of the software. Did it change more than $X\%$?
\end{enumerate}

This system was a vast improvement over just staring at graphs. It was automated and computers do not
get bored or lose focus. However, we soon realized we had a large number of false positives with
which to deal (up to 99\% as discussed in Section~\ref{sec:impact} depending on how you count). All the tests have some inherent noise in them. However, different tests and
different test configurations have different levels and types of noise. Using a static threshold
(e.g., $10\%$) led to a lot of alerts for changes on noisy tests. At the same time it would either miss
real changes that were less than the threshold, or detect the smaller changes at some later time (when the change plus
the noise crossed the threshold). For
example, in Figure~\ref{fig:trend-map} the insert\_ttl test shows a clear, but very small
change around June 13th, while the map\_reduce workload is steady with run to run variation (noise). There is no way we could adjust a common threshold for those two tests that
would both catch the change in insert\_ttl and not constantly flag the map\_reduce test, or constantly flag the
insert\_ttl 1 thread line after July 19th.

While the system was noisy and painful, it was still useful for us. Over time we added fixes and
band aids to the system to reduce the number of false positives (e.g., adjust the thresholds
separately for different tests). Even with all those fixes to the previous system, the system:
\begin{itemize}
\item Produced a large number of false positive alerts that had to be processed
\item Missed smaller regressions (false negatives)
\item Flagged failures on commits unrelated to the regression, requiring humans to trace back to find when a regression actually happened
\item Only flagged regressions, not improvements
\item Consumed a huge amount of time from the team to process all the results
\end{itemize}

We wanted to improve the process to something fundamentally better. We started by trying to clearly
state the problem.

\section{Problem Definition}\label{sec:problem}

The software we are testing is updated in discrete commits to our source repository. Our problem is to:
\begin{problem}
  Detect which commits change the performance of the software (as measured by our performance tests)
  in the presence of the noise from the testing infrastructure.
\end{problem}

When the problem is phrased like that, it is clear that this is a signal processing problem. The
noise of the system makes the performance results a stochastic process. As such, we reviewed the
literature on signal processing\footnote{See Section~\ref{sec:related} for references on signal
  processing and change point detection}.  In particular, we decided our problem was a change point
detection problem. From Matteson and James~\cite{edivisive2014}: ``Change point analysis is the process
of detecting distributional changes within time-ordered observations.'' Our
system can be represented as the time series:
\begin{equation}
  \label{eq:signal}
S_t = P_t + N_t
\end{equation}
Where $S_t$ is the measured signal, $P_t$ is a constant value of performance, and $N_t$ is a random variable
representing the noise. It is not clear a priori that the noise component is independent,
identically distributed (IID) or not. In fact, we have examples in which the noise is correlated with
time.

The noise is really composed of at least three components:
\begin{equation}
\label{eq:noise}
N = N_p + N_s + N_w
\end{equation}

\begin{itemize}
\item $N_p$: The noise from the test platform, including CPU, network, OS, etc.
\item $N_s$: The noise from the software itself. This can be from the compiler as well as
    other non-determinism or randomness in the software.
\item $N_w$: The noise from the workload generator. The workload generator is also software and may
  vary in the load it applies to the system.
\end{itemize}
Reducing $N_p$ and $N_w$ was the focus of previous work~\cite{Ingo2019}. However, as computers and
computer networks are inherently very complicated, non-deterministic machines, including things like
caches and predictors, we can only hope to reduce those noise components, not remove them. As such,
we have to learn to live with noise.

\subsection{E-Divisive Means}\label{subsec:edivisive}

After reviewing the literature on change point detection, we decided to use the E-Divisive means
algorithm~\cite{edivisive2014}. E-Divisive means has the following useful properties for us:
\begin{itemize}
\item Does not require any distributional assumptions other than a finite mean
\item Can be recursively applied to identify multiple change points
\item Does not require training data and works out of the box on a time series
\item Optimal sets of change points can be efficiently computed via dynamic programming and permutation sampling
\end{itemize}
Additionally, the main focus of our work is on retroactive identification of change points, not
prediction. Many time series algorithms are intended for forecasting, which is not relevant in the context of commits
being pushed by developers.

The algorithm works by hierarchically selecting distributional change points that divide the time
series into clusters. For a given cluster, a candidate change point is chosen by computing a
statistic $\widehat{Q}$ at every point within the cluster and selecting the point with the largest value.
 This
statistic $\widehat{Q}$ is the discrete analog of the divergence between continuous multivariate
distributions. After the first change point is determined, the algorithm is then reapplied to each
cluster created by the existing change point. This process repeats until the largest candidate $\widehat{Q}$ value
is below a threshold of significance.

\subsubsection{$\widehat{Q}$ Statistic Computation}\label{subsec:qhat}
To compute $\widehat{Q}$ for a series of n + m points bisected at the nth
point, let:

\begin{equation}
  \label{}
\mathbf{X}_n = \{ X_i: i=1, ..., n\}
\end{equation}
\begin{equation}
  \label{}
  \mathbf{Y}_m = \{ Y_j: j=1, ..., m\}
\end{equation}

Represent the first n points and the last m points, respectively.  An empirical
divergence metric can be computed as follows:

\begin{equation}
  \label{eq:ehat}
  \begin{split}
  \widehat{\mathcal{E}}(\mathbf{X}_n, \mathbf{Y}_m;\alpha) = & \frac{2}{mn}\sum_{i=1}^n\sum_{j=1}^m{|X_i - Y_i|}^\alpha
  \\ & - \binom{n}{2}^{-1}\sum_{1\leq i < k \leq n}{|X_i - X_k|}^\alpha \\ & - \binom{m}{2}^{-1}\sum_{1\leq j < k \leq m}{|Y_j - Y_k|}^\alpha
\end{split}
\end{equation}
The $\alpha$ parameter can be any value between 0 and 2. We use 1 for simplicity\footnote{Matteson
  and James~\cite{edvisive2014} observed similar results with $\alpha$ of 0.5 and 1.5, and presented their simulation studies with
  $\alpha$ of 1.}.
We then attain $\widehat{Q}$ by weighting the previous result by the size of our clusters:

\begin{equation}\label{eq:qhat}
\widehat{Q}(\mathbf{X}_n, \mathbf{Y}_m;\alpha) = \frac{mn}{m+n}\widehat{\mathcal{E}}(\mathbf{X}_n, \mathbf{Y}_m;\alpha)
\end{equation}

\subsubsection{Termination}\label{subsec:label}

The procedure for termination in the paper involves numerous random permutations of the values in
each cluster to determine the statistical significance. Intuitively, if rearranging the order of data
in a cluster does not significantly affect the maximum $\widehat{Q}$, there is likely no meaningful
change point.

\section{Implementation}

Selecting an algorithm for detecting change points was just the beginning. We then needed to:
\begin{itemize}
\item Implement (or find an implementation of the algorithm)
\item Automate processing our data with the algorithm
\item Present the data after processing
\item Enable users to act on the presented data
\end{itemize}

The first step of implementing the algorithm was straightforward as there is an R
package~\cite{rpackage2007} for the algorithm. We also wrote a Python implementation\footnote{The
  Python implementation of e-divisive means is available here:
  https://github.com/mongodb/signal-processing-algorithms} to go with our existing Python code
base. Both implementations will iteratively find the most significant change point in the series,
and then analyze the clusters on either side of the change point, until reaching the stopping
criterion. Even though we used our Python implementation, we benefited from the existence of the R
implementation. We were able to compare the results from the two implementations to gain confidence
in the correctness of our implementation. It also gave us a reference point for the performance of
the algorithm.

It took a little more work to transform our existing data into a form that could be consumed by the
algorithm. We store results in a MongoDB database, with all the results for a given task build
stored together. We unbundled the data so we could easily generate a vector of results for each
individual combination of (system variant, task, test, test configurations). We then saved the
computed change points in the database in a separate
collection. Each change point has the same identifying fields as the results data, plus data from
the algorithm itself, such as the $\widehat{Q}$ and the order of the change point (was it the first
or $N^{th}$ change point for the series when recursively identifying change points). After
identifying the change points for a series, each series is partitioned into clusters in which the
results are unchanging. We call those clusters ``stable regions''. We
calculate common statistics for each stable region, and include that in the entry for each change
point. Statistics include the min, max, median, mean, variance for the given measure, as well as how
many data points are in the stable region.

The algorithm is run at the completion of every set of performance tests. The result data is
uploaded and transformed before the algorithm is run. We replace any existing change points with
the newly generated change points. Rerunning the analysis allows the change points to be updated as
more data becomes available. The data is statistical in nature, so having more data allows better
analysis. As such, it is expected that on occasion a previously computed change point will ``move''
to a new commit or even completely go away as more data becomes available. Additionally, the change
point algorithm will never find that newest result to be a change point. If a regression was
introduced, the algorithm will only detect a change point after several more test results are
collected. A three to five day delay is common for our tests. This contrasts with processing results for
correctness tests in which the pass/fail result is determined at run time and will not change at a
later time.

\subsection{Display}

\begin{figure}[ht]
  \centering
  \includegraphics[width=\linewidth]{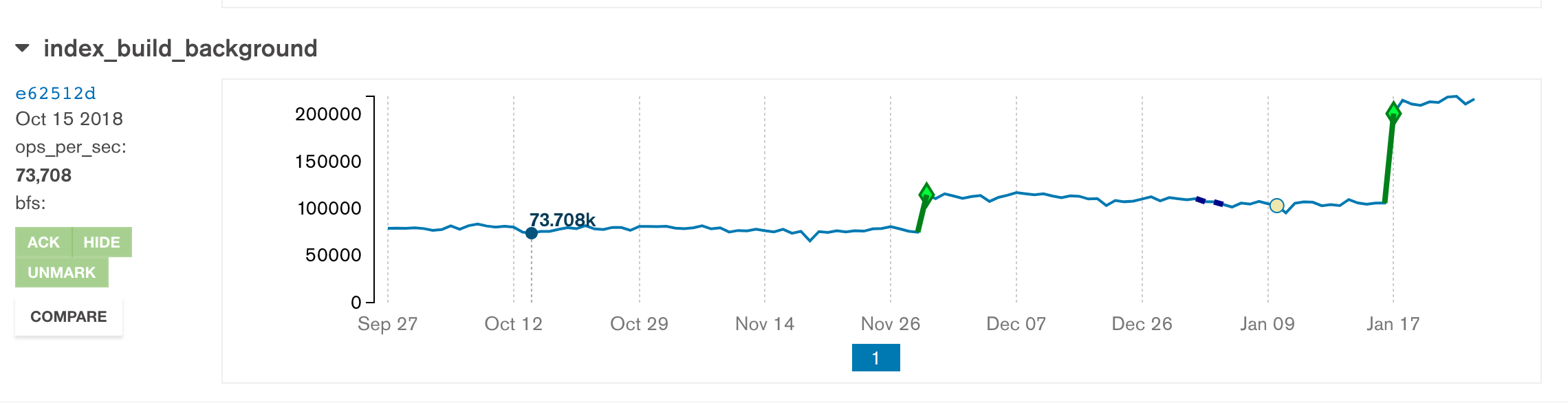}
  \caption{\label{fig:trend-index} Trend graph for index build background test. Notice two change
    points (green highlights) and two JIRA tickets (diamonds).}
\end{figure}

\begin{figure}[ht]
  \centering
  \includegraphics[width=\linewidth]{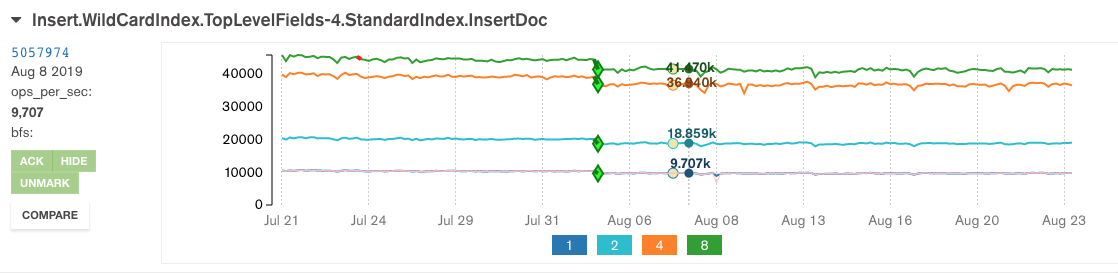}
  \caption{\label{fig:trend-wildcardindex} Trend graph for the wildcard index test with annotations
    for Jira tickets and change points. This figure can be compared to
    Figure~\ref{fig:wildcard-clean} which does not have the annotations.}
\end{figure}

We then displayed the data in two distinct ways. First, we updated the trend graphs discussed in
Section~\ref{subsec:existing}. We added annotations to show the change points as well as any Jira
tickets that match the test and revision. Figures~\ref{fig:trend-index}
and~\ref{fig:trend-wildcardindex} show two trend graphs annotated with both change points and Jira
tickets.  The green diamonds are the Jira tickets, and the highlighted segments are change
points. You can clearly see two tickets matching two green highlighted sections in
Figure~\ref{fig:trend-index}, while there is a small drop with change point and ticket highlighted
in Figure~\ref{fig:trend-wildcardindex}. That figure is particularly interesting as the change in
mean is on the same scale as the level of the noise before and after the change, and a version of
that graph without annotations appeared earlier in Figure~\ref{fig:wildcard-clean}. As we will discuss
in Section~\ref{sec:triage}, it is a goal of our process that each change point is matched to a Jira
ticket or hidden from view if it is clear noise.

We added necessary meta data fields to our Jira tickets so that they can be mapped to their
originating Evergreen project, task, test and git revisions. We track both ``First failing
revision'' and ``Fix revision''. Additionally, we needed to query this data every time we drew the
displays. For performance reasons, we setup a regular batch job to query Jira and load the data into
our database.

Note that we do not run the performance tests on every commit to the source repository. That adds
complexity to the visualization above. A change point is identified for commits that were built and
run. The algorithm cannot determine which commit caused a performance change if the commits have not
run. Instead, we can identify the range of commits between the change point and the previous test result as suspect commits. When displaying these results and
annotations, we must be able to match up change point, Jira ticket, and builds that have actually
run, even if they point to close but different commits.

\begin{figure*}[ht]
\centering
\includegraphics[width=\linewidth]{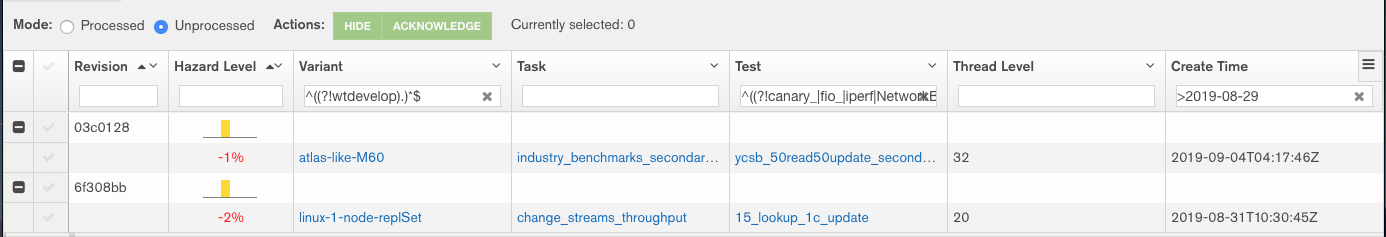}
\caption{\label{fig:bb-unprocessed} Triage view of unprocessed change points. }
\end{figure*}

\begin{figure*}[ht]
\centering
\includegraphics[width=\linewidth]{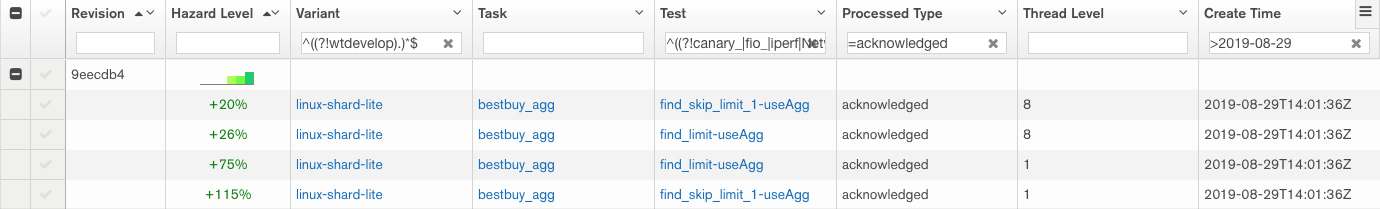}
\caption{\label{fig:bb-processed} View of processed change points\protect\footnotemark.}
\end{figure*}

Those annotations are great for examining how one test performs over time, and identifying the
location of significant changes. It is not useful though for tracking all tests, triaging the most
important regressions, and grouping tests that change at the same time. For that, we built a new
page for viewing the change points. This new page shows all the change points on one page, grouped
by commit and is shown in Figures~\ref{fig:bb-unprocessed} and~\ref{fig:bb-processed}. Change points
are initially ``unprocessed'' and through triage become ``processed''.

Figure~\ref{fig:bb-unprocessed} shows the list of software revisions (identified by their git hashes) associated with ``unprocessed'' change points. All the
change points for a single git hash are grouped together. The display has the git hash, test name,
task name, date of the build, and the hazard level of the change. Hazard level is computed as the
log of the ratio of mean before to the mean after the change point. The hazard level has proven
extremely useful in focusing triage on the major regressions and it has the property that two
changes that cancel out (e.g., $33\%$ drop followed by a $50\%$ rise) have the same magnitude. The display has the ability to filter or sort
on any of the fields. There are two
buttons along the top to acknowledge or hide the change points. Hiding or acknowledging makes a copy
of the changed point with the addition of field to indicate that the change point has been
processed. The process of acknowledging or hiding a change point also removes it from the
unprocessed change point page. It will also be removed from this page if there is already a Jira
ticket for the same test at that commit. The list is short in the figure because the remaining change points had
already been ``processed'', with only two less interesting change points remaining.

\footnotetext{Our Best Buy tests are  based off of the open data set from Best Buy~\cite{bestbuy}}
Once a commit is acknowledged or hidden, it shows up on the processed tab as shown in
Figure~\ref{fig:bb-processed}. There we have shown one change point with a significant improvement in
performance. We will isolate that change to a single commit, and open a Jira ticket to document
it. Once that ticket is opened, the change point will also be removed from this view.

\subsection{Triage}\label{sec:triage}

We have a rotating role of a dedicated person to check for performance changes. We call this person
the ``Build Baron''. The Build Baron uses all the views of the change points shown in
Figures~\ref{fig:bb-unprocessed} and~\ref{fig:bb-processed} along with the trend graphs as shown in
Figures~\ref{fig:trend-index} and~\ref{fig:trend-wildcardindex}. Ideally, the Build Baron reviews
every change point, deciding if it is a real change worthy of investigation, if it is insignificant,
or if it is just noise.

The Build Baron starts with the list of the unprocessed change points as shown in
Figure~\ref{fig:bb-unprocessed}. By default we have the list set to filter out some things (lower
priority, canary\footnote{Canary tests are tests run to detect changes to the test system
  itself. Changes in performance for canary tests indicate a problem with the test
  itself. See~\cite{Ingo2019} for more on our use of canary tests.} tests, or some results that are
merely informational). The Build Baron goes down this list investigating each git hash in the
list. Each git hash can have multiple change points (multiples tests, variants, etc). The test
column includes links to the task page with the matching trend graph for the Build Baron to decide
if the change point is worth further investigation.

They may determine from inspection that there is a clear change. If so, they will try to isolate the
offending commit. Recall that we do not run every commit, so once a performance change is detected,
we need to run the tests on the missing commits to determine the precise commit that introduced the
performance change. The Build Baron will schedule the builds appropriately, and ``acknowledge'' the
change point by pressing the ``acknowledge'' button on the change point page. The Build Baron will
need to wait for the tests to finish. When the tests finish, the system will rerun the change point
detection algorithm in order to update the change point to the correct commit. If the change point
represents a new issue, the Build Baron will open a Jira ticket for the change and assign it to the
appropriate team to investigate. If there is an existing Jira ticket (e.g., this is the fix to a
previous regression) the Build Baron will update the existing Jira ticket. If the performance change
is a significant enough regression, we may also revert the commit.

Not all change points lead to Jira tickets. Occasionally we will have spurious drops in performance,
with the performance changing for one build and then returning to previous performance. Drops such
as these may go away if we rerun the task. If the performance change is large enough it can lead to
a change point being created. We start by checking these against our canary workloads.
As part of previous noise reduction work we have a number of canary workloads. The canary
workloads tell us nothing about the software we are testing, but a lot about the test system. The
results should be constant. If we detect a statistically significant change for a canary or
otherwise suspect a spurious drop, we treat
all the results from that test run as suspect and rerun the test.

Alternatively, the change point may
be due to noise. The algorithm is very good at accounting for IID noise, but not for correlated
noise. Unfortunately, we have some correlated noise in our system -- some from time varying behavior
on our cloud providers, other due to things as finicky as code alignment issues in compilation and
its impact on processor caches. Either case usually can be confirmed visually, and the
Build Baron will ``Hide'' the change point by using the Hide button. That will remove the change
point from the change point page and change the color of the change point on the trend graph to blue. If we
have a sustained change in performance due to cloud changes, we open a Jira ticket
for that in order to document it. That way, anyone looking at the trend graphs will be able to see
why the performance changed.

All change points that are processed (hidden or acknowledged) and all change points that match a Jira ticket are removed from
the change point list. The list is meant for triage and those points have already been triaged. We
have the history of the points however and can use it to compile effectiveness data about our
system.

Since we group change points by git hash in the display, we may have the
case in which we want to acknowledge some of the change points in the group and hide others. This
is fully supported.

\section{Optimizations}\label{sec:optimizations}

The computational complexity of a naive implementation of E-Divisive means is
$\mathcal{O}(\kappa T^3)$, where $\kappa$ is the number of change points and $T$ the number of
observations in the series. Matteson and James~\cite{edivisive2014} point out
that it is possible to compute the series $\widehat{Q}$ in $\mathcal{O}(T^2)$ (per change point) as
each value in the series can be derived from the previous with linear time. We implemented this
optimization in our POC.

The major generations of the implementation were:
\begin{itemize}
\item Naive $\mathcal{O}(\kappa T^3)$ implementation
\item $\mathcal{O}(\kappa T^2)$ implementation (row/column differences)
\item Use NumPy arrays
\item Use native C implementation (Python module)
\end{itemize}
We also tested compiling the first 2 versions with Cython, but this yielded only about $10\%$
improvements and was discarded as an option.

With a test data series of 173 values, using an Intel(R) Core(TM) i7-6560U CPU @ 2.20GHz, single
threaded executions yielded the following results:
\begin{center}
\begin{tabular}{ |c|c|c| }
 \hline
 Implementation & Time (s) & Ratio \\
 \hline
  Naive  & 0.622551 & 6103.44 \\
  Row/column differences  & 0.011567 & 113.40 \\
  NumPy              & 0.001282 & 12.57 \\
  Native             & 0.000102 & 1.00 \\
 \hline
\end{tabular}
\end{center}
At this modest problem size, the native C implementation is over 6,000x faster than our original
naive implementation, 113x faster than a non-naive implementation, and 13x faster than our best Python
only implementation.

At the end of a CI task there are 20-60 new results. The historical time series for those is fetched
and the algorithm is used to re-compute change points. The computation is easy to parallelize as
there are multiple series to be processed. Using the native C implementation and parallelizing the
computation, we are able to re-compute all the change points in under 10 seconds. The majority of
that time is spent waiting for network communication to the database, with the E-Divisive mean
computation taking only milliseconds. The computation performance is a combination of the
performance enhancements discussed above, and a conscious decision to limit our analysis to results
from N recent months, by using up to 500 data points\footnote{This is approximate. If more than 500 data points are
  available, we will select data points
  going back to an already computed change beyond that range. As such, the computation will use more
than 500 points.}. We do this since we are interested in catching regressions in new commits
and new releases. Due to this, our time series typically have 100 to 500 values and 1 to 10 change
points, and the computation time is minimal.

In our original project plan we were concerned about the $\mathcal{O}(\kappa T^3)$ complexity and
had envisioned running change point detection in a separate asynchronous job so as to not impact
test execution time. Due to the above optimizations we are able to run the change point detection
immediately after each test execution in a simpler process. When the Build Baron is
running additional tests to isolate the exact commit that changed performance, having the change
point detection immediately refreshed reduces the time to isolate the change.

Even though the computation time is minimal in production, we still care about the execution
time. Occasionally we need (or want) to recompute all the change points in our system (e.g., we
updated something related to the computation or we fixed a bug). Based on the current performance of
the algorithm, we believe we could recompute all the change points across our system in under an
hour. Previously we had a need to regenerate change points for a section of the data and we tried to
regenerate the points over the weekend. The job was still running Monday morning when we checked on
it. When we tried to regenerate the data more recently for the same subset, it took approximately 30
minutes. Finally, we may want to extend the length of history we use in the computation in the
future, for example if we add a project that runs much more frequently.

\section{Impact}\label{sec:impact}

The point of this work, as discussed in Section~\ref{sec:problem}, was to more efficiently detect
which commits change the performance of our software. It has succeeded. Triaging performance changes
has gone from being a full time job that one person could not completely keep up with, to one that
the Build Baron can process all of the change points and have time left over. There is a morale improvement, as the Build Baron is working on real performance
changes that make a difference to the company. Additionally, we are able to detect smaller changes in
performance than we could previously.

We are also tracking performance improvements now in addition to performance regressions. This has
had a few impacts:
\begin{itemize}
\item We have caught correctness bugs (e.g., important check suddenly skipped).
\item We have been able to provide positive feedback to developers for making things faster. Someone
  notices the impact of their work now.
\item We can share improvements with marketing~\cite{DJ2019}.
\end{itemize}

While we clearly and strongly know that we have qualitatively improved our system, we have also
quantitatively improved it. We have two sets of data: First, we performed a proof of concept (POC)
before fully implementing the system. During the POC we looked at false negative and false positive
rates, as well as general feasibility related issues. Since then we have implemented the system in
full, including collecting the labeling data and the information in Jira tickets. The combination
of that data allows us also to measure the effectiveness of the system in production.

\subsection{Learnings from the POC}\label{subsec:POC}

For the POC we implemented the calculation of change points as a batch job. We then
compared the changes points to the results from our existing system and in Jira tickets. We focused
on a 5 month period of results. At that time our automated system used two level of Jira
tickets. The first level was automatically generated by the analysis code. The Build Baron then
would go through that list. The Build Baron would move representative failures to the second Jira
project, and link the rest. For the 5 month period we had 2393 tickets in the first project, and 160
in the second. Of the 160 tickets in the second project, 24 of them were ``useful.''\footnote{For
  this use we  considered a ticket useful if it reflected a change in performance of the
  software.} So, on average 100
tickets in the first project reduced down to 1 useful ticket in the second project, with a human
going through all of those. In other words, the old system generated a huge number of Jira tickets,
of which the vast majority were false positives, while also suffering from false negatives.

The change point algorithm found all the regressions we were tracking in Jira tickets. The change point algorithm
may have missed performance changes, but if it did those were performance changes we were already
missing. From a false negative rate, the new system was strictly better than the old system.

We also looked at the change points that did not match existing tickets. We started by sorting on the
$\widehat{Q}$ value. The first three that we investigated led to two new Jira tickets. The first
element was one ticket, while the second and third items shared a ticket. The second item
corresponded to a drop in performance and and the third item was a later fix for the earlier drop.

Since we did not have complete reference data for all the performance data, we instead sampled the
change points to estimate a false positive rate. We considered only the change points that did not
match existing Jira tickets, as those are the potential false positives. We looked at 10 of those
change points. Of those:
\begin{itemize}
\item 2 were clear changes in performance we had not caught previously.
\item 4 were clearly noise.
\item 4 were not clear cut. They included
  \begin{itemize}
  \item One clear change in distribution (the test got noisier)
    \item Two cases in which the distribution changed slightly. We think this was likely correlated noise in
      our system.
    \item One large outlier value got the system to create a change point.
  \end{itemize}
\end{itemize}

Of the cases that were noise we used the order of change point statistic to search for more
significant regressions. In all cases there was a point in which change points went from noise or maybe
regressions, to clear cut changes.

Depending on how you count the four ``not clear cut'' cases, we had a false positive rate between 40\% and
80\%, with the possibility to tune the system. Given the previous system required looking at 100
things for each useful one, this was a huge improvement and we knew we wanted to move forward.

\subsection{Learnings from Production}\label{subsec:production}

After the POC we put the change point detection algorithm into production and have been collecting
data. Looking at a 3 month period from summer of 2019 for our primary performance project we have:
\begin{itemize}
\item 12,321 distinct change points
\item Corresponding to 178 unique git hash values
\item 126 of those 178 correspond to a Jira ticket
\item 122 of those 178 were acknowledged or hidden
\item there were 79 unique Jira tickets for those change points
\end{itemize}

Therefore, the Build Baron had 178 things to look at during this period, and more than 2/3 of all of
them were issues we felt were worth tracking and possibly working on. We went from 100 notifications
leading to 1 tracked issue, to 9 notifications leading to 6 worth follow-up and 4 tracked
issues. Given that we want the process to be slightly biased toward false positives, this result is
close to optimal.

The process still is not perfect though. We
also have:
\begin{itemize}
\item Acknowledged change points that do not match a raw change point
\item Acknowledged change points without a matching Jira ticket
\item Hidden change points associated with Jira tickets
\end{itemize}

The two cases of acknowledged change points not matching a Jira ticket or a raw change point appear
to come from a few cases related to triaging: aggressive triaging and mistaken triaging. The change
points are grouped by git hash. It is easy to check some of the change points for the git hash and
observe a distinct performance change, and aggressively acknowledge all the points. Alternatively,
sometimes someone hits the wrong button and acknowledges when they probably should have hidden a
point. We observe both of those cases when inspecting points for
these two cases. We are adding more auditing of how we label change points.

Hidden change points that do not match a raw change point mainly come from a combination of marginal change
points aggressively triaged, in which the change point algorithm is able to exclude once it has
additional data.

\section{Open Problems}\label{sec:openproblems}

This work is part of a larger effort to understand the performance of our software, and to know when
the fundamental performance changes. This builds on work to
\begin{itemize}
\item Build tools for making performance workloads
\item Automate running of performance tests
\item Include external benchmarks in our system
\item Reduce the noise in our test bed
\end{itemize}

There are a number of areas we continue to work on in this space
\begin{itemize}
\item Better workload generation tools
\item More flexible result types
\item Better signaling of performance changes
\item Determine if a ``patch'' has changed performance
\item Identify, exclude, and rerun test runs in which ``canaries'' have failed
\item Use the data from processing change points to make the algorithm more effective
\end{itemize}

Better workload generation tools is about making it easier for our engineers to craft interesting
workloads, while the more flexible result types deal with handling latencies, percentiles,
throughputs, etc.

\begin{figure}[ht]
  \centering
  \includegraphics[width=\columnwidth]{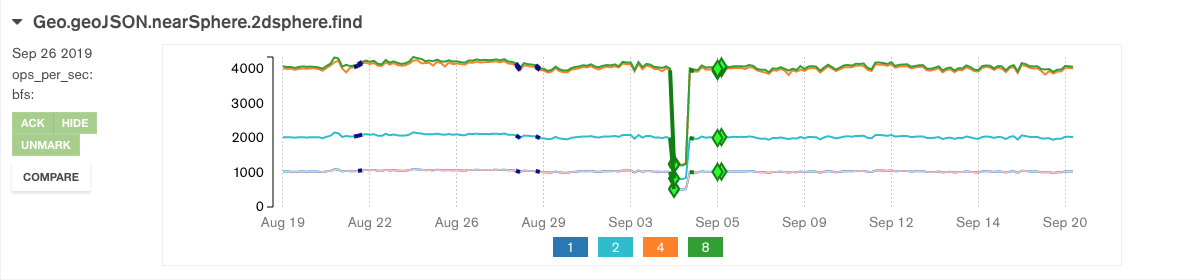}
  \caption{\label{fig:quickdrop} Example of a drop in performance followed quickly by a fix. The
    regression and fix are too close together for the E-Divisive means algorithm to identify both changes. }
\end{figure}

The next three items show the limits of the change point detection algorithm. E-Divisive means is
great at detecting changes in distribution, but it needs multiple data points to compute that
distribution. It also has a bias towards the center of the longest clusters. As such, it cannot tell you anything about the most recent test run, and it will not
detect two distinct changes if they are too close together, such as in
Figure~\ref{fig:quickdrop}. In Figure~\ref{fig:quickdrop} a performance impacting code change went
in and was reverted soon afterwards. There are not enough commits for the algorithm to detect that
two distinct changes happened.

The bias towards the center of clusters is a result of the coefficient $mn / (m + n)$ in
Equation~\ref{eq:qhat}. The coefficient was an integral assumption in the consistency proof
in~\cite{edivisive2014}. Since $m + n = T$, where $T$ is the number of observations in the series,
we get $mn/(m+n) = m(T-m)/T = m - m2/T$. The first and last values of this coefficient are therefore
$1 - 1/T$ when either $m$ or $n$ equals $1$. All other values of this coefficient are greater than
one (e.g. $2 - 4/T$ for $m = 2$) and the coefficient achieves the maximum value of $T/4$ when
$m=n=T/2$. Even if it would lead to possibly invalidating the mathematical proof behind E-Divisive,
in practice it is tempting to explore whether these limitations could be removed.

The algorithm has a parameter for the minimum cluster size of observations: N. The tests used later
in~\cite{edivisive2014} used minimum cluster sizes N=30 and N=10, while we have operated with N=3 to
have faster notification. With N=3 we require at least 3 data points after the change (including the
change) to detect the change. The combined effect of these limitations is that in our primary
performance project we typically get alerts of a new regression 5-6 days after it was committed.

Ultimately change point detection is not suitable for determining whether a single new test result
is a regression or not. We can use outlier detection to answer that question. The change point
detection algorithm can identify a stable region, and then we can use outlier detection to
determine if the patch build is statistically the same as the stable region or not. This will allow
us to react early to very clear regressions, while change point detection can catch smaller
regressions after a delay. We also have a use case in which developers can test (patch test)
their code changes (patches) before pushing them to the source repository. We can use the same
outlier detection to detect performance changes in patch tests. Finally, with our canary tests the goal is reversed:
Rather than ignoring outliers, we want to be alerted about outliers so that we can rerun the tests
or quarantine the results.

As mentioned in Section~\ref{sec:triage}, we run a number of canary tests to detect changes to the
test system itself. If we have a sudden change in the canary workloads we would like to flag the
results and possibly rerun the tests. The detection of performance chances in the canary workloads,
flagging the results, and rerunning suspect executions are all amenable to automation.

Finally, we are collecting a large corpus of training data related to the change point algorithm,
through the acknowledged and hidden change points. This will allow us to compare the existing
algorithm to other change point detection algorithms. Additionally, we expect we can use this data to
either manually adjust the algorithm, or train more advanced machine learning algorithms.

\section{Related Work}\label{sec:related}

Previous work has focused on reducing the noise in our test environment~\cite{Ingo2019}. Our test
environment is built on the Evergreen~\cite{Erf2016} CI system. There are many other continuous
integration systems, such as Jenkins~\cite{smart2011jenkins}, BuildBot~\cite{buildbot}, Travis
CI~\cite{TravisCI}, Gitlab CI~\cite{GitlabCI}, and Circle CI~\cite{CircleCI}. The other CI systems
have various supports for graphing performance results. For instance, Jenkins has a performance
plugin~\cite{JenkinsPerformance} to support running common benchmarking tools, reporting results,
and plotting trend graphs.

The Chromium project has implemented its own CI system called LUCI~\cite{luci} and implemented regression
detection~\cite{ChromiumTesting} within it. Each test requires setting explicit threshold values for alerting a
regression. As such, it is very similar in concept to our previous analysis system.

Google has a patent filing~\cite{GoogleWindowDetection} for an algorithm called
``Window deviation analyzer'' (WDA). It describes identifying windows of data in a time
series, computing median and mean for each window, and estimating if the most recent
sample differs more than some threshold from those values. Like our system, Google's WDA
will fail a test some number of commits after the commit that introduced a regression. Unlike our
system, WDA requires an explicit threshold and it is does not identify a specific
offending commit (it identifies a window of commits within
which a regression has occurred).

For algorithms for dealing with noise in time series data, we found the NIST/Sematech e-handbook on
Statistical Methods~\cite{nisthandbook} a good overview, particularly chapter 6 ``Process and
Product Monitoring and Control''. We looked into using techniques such as Moving Averages or Box
Jenkins~\cite{box2015time} for the time series data. That would involve building a model of the
process and detecting when samples diverge from the existing model.

Additionally, see~\cite{cp2017survey} for an extensive survey on change point detection techniques
in time series data, covering supervised methods such as multi-class, binary, and virtual
classifiers, and unsupervised methods such as likelihood ratio, subspace models, and probabilistic
models. There are frequentist/Bayesian and parametric/non-parametric methods algorithms for both the
supervised and unsupervised classes.

\section{Conclusion}

In this paper we describe how we detect performance changes in our software using our continuous
integration system and change point detection. We run a large collection of tests periodically as
changes to our software product are committed to our source repository. We would like to know when
the performance changes for those tests. Originally we used humans looking and graphs, which was
later replaced with a threshold based automatic detection system, but neither system was sufficient
for finding changes in performance in a timely manner. This paper describes our move to using change
point detection (specifically E-Divisive means) to detect and notify when the performance changes.

In addition to implementing the change point algorithm, we also had to integrate it into our
existing performance testing system and visualize the results so that engineers could triage and
use the information. We describe what was required to get change point detection working in our
system and how we triage and process the results. We also describe our efforts to speed up the
algorithm so that we could use it synchronously within our workflow.

The net impact of this work was large for us: Quantitatively, it dramatically dropped our false
positive rate for performance changes, while qualitatively it made the entire process easier, more
productive (ex. catching smaller regressions), and more timely.

There is more work to be done to continue to improve our ability to detect when the performance of
our software changes. We discuss a number of open problems for performance testing in general and
for using change point detection in particular.

 \bibliographystyle{ACM-Reference-Format}
 \bibliography{bibfile}


\begin{thebibliography}{19}


\ifx \showCODEN    \undefined \def \showCODEN     #1{\unskip}     \fi
\ifx \showDOI      \undefined \def \showDOI       #1{#1}\fi
\ifx \showISBNx    \undefined \def \showISBNx     #1{\unskip}     \fi
\ifx \showISBNxiii \undefined \def \showISBNxiii  #1{\unskip}     \fi
\ifx \showISSN     \undefined \def \showISSN      #1{\unskip}     \fi
\ifx \showLCCN     \undefined \def \showLCCN      #1{\unskip}     \fi
\ifx \shownote     \undefined \def \shownote      #1{#1}          \fi
\ifx \showarticletitle \undefined \def \showarticletitle #1{#1}   \fi
\ifx \showURL      \undefined \def \showURL       {\relax}        \fi
\providecommand\bibfield[2]{#2}
\providecommand\bibinfo[2]{#2}
\providecommand\natexlab[1]{#1}
\providecommand\showeprint[2][]{arXiv:#2}

\bibitem[\protect\citeauthoryear{??}{bes}{[n.d.]}]%
        {bestbuy}
 \bibinfo{year}{[n.d.]}\natexlab{}.
\newblock \bibinfo{title}{Best Buy APIs open data set}.
\newblock
\newblock
\urldef\tempurl%
\url{https://bestbuyapis.github.io/api-documentation/#overview}
\showURL{%
\tempurl}


\bibitem[\protect\citeauthoryear{??}{Chr}{[n.d.]}]%
        {ChromiumTesting}
 \bibinfo{year}{[n.d.]}\natexlab{}.
\newblock \bibinfo{title}{(Chromium) Regression Detection for Performance
  Tests}.
\newblock \bibinfo{howpublished}{wiki}.
\newblock
\urldef\tempurl%
\url{https://www.chromium.org/chromium-os/testing/perf-regression-detection}
\showURL{%
\tempurl}


\bibitem[\protect\citeauthoryear{??}{Cir}{[n.d.]}]%
        {CircleCI}
 \bibinfo{year}{[n.d.]}\natexlab{}.
\newblock \bibinfo{title}{Circle CI}.
\newblock
\newblock
\urldef\tempurl%
\url{https://circleci.com/docs/}
\showURL{%
\tempurl}


\bibitem[\protect\citeauthoryear{??}{Eve}{[n.d.]}]%
        {Evergreen}
 \bibinfo{year}{[n.d.]}\natexlab{}.
\newblock \bibinfo{title}{Evergreen Continuous Integration}.
\newblock
\newblock
\urldef\tempurl%
\url{https://github.com/evergreen-ci/evergreen/wiki}
\showURL{%
\tempurl}


\bibitem[\protect\citeauthoryear{??}{Git}{[n.d.]}]%
        {GitlabCI}
 \bibinfo{year}{[n.d.]}\natexlab{}.
\newblock \bibinfo{title}{Gitlab CI/CD}.
\newblock
\newblock
\urldef\tempurl%
\url{https://docs.gitlab.com/ee/ci/}
\showURL{%
\tempurl}


\bibitem[\protect\citeauthoryear{??}{Jen}{[n.d.]}]%
        {JenkinsPerformance}
 \bibinfo{year}{[n.d.]}\natexlab{}.
\newblock \bibinfo{title}{Jenkins Performance Plugin}.
\newblock \bibinfo{howpublished}{wiki}.
\newblock
\urldef\tempurl%
\url{https://wiki.jenkins.io/display/JENKINS/Performance+Plugin}
\showURL{%
\tempurl}


\bibitem[\protect\citeauthoryear{??}{luc}{[n.d.]}]%
        {luci}
 \bibinfo{year}{[n.d.]}\natexlab{}.
\newblock \bibinfo{title}{(LUCI) A Tour of Continuous Integration UI}.
\newblock
\newblock
\urldef\tempurl%
\url{https://chromium.googlesource.com/chromium/src.git/+/master/docs/tour_of_luci_ui.md}
\showURL{%
\tempurl}


\bibitem[\protect\citeauthoryear{??}{Tra}{[n.d.]}]%
        {TravisCI}
 \bibinfo{year}{[n.d.]}\natexlab{}.
\newblock \bibinfo{title}{Travis CI}.
\newblock
\newblock
\urldef\tempurl%
\url{https://docs.travis-ci.com}
\showURL{%
\tempurl}


\bibitem[\protect\citeauthoryear{Ablett, Sharlin, Maurer, Denzinger, and
  Schock}{Ablett et~al\mbox{.}}{2007}]%
        {buildbot}
\bibfield{author}{\bibinfo{person}{Ruth Ablett}, \bibinfo{person}{Ehud
  Sharlin}, \bibinfo{person}{Frank Maurer}, \bibinfo{person}{Jorg Denzinger},
  {and} \bibinfo{person}{Craig Schock}.} \bibinfo{year}{2007}\natexlab{}.
\newblock \showarticletitle{Buildbot: Robotic monitoring of agile software
  development teams}. In \bibinfo{booktitle}{\emph{RO-MAN 2007-The 16th IEEE
  International Symposium on Robot and Human Interactive Communication}}. IEEE,
  \bibinfo{pages}{931--936}.
\newblock


\bibitem[\protect\citeauthoryear{Aminikhanghahi and Cook}{Aminikhanghahi and
  Cook}{2017}]%
        {cp2017survey}
\bibfield{author}{\bibinfo{person}{Samaneh Aminikhanghahi} {and}
  \bibinfo{person}{Diane~J Cook}.} \bibinfo{year}{2017}\natexlab{}.
\newblock \showarticletitle{A survey of methods for time series change point
  detection}.
\newblock \bibinfo{journal}{\emph{Knowledge and information systems}}
  \bibinfo{volume}{51}, \bibinfo{number}{2} (\bibinfo{year}{2017}),
  \bibinfo{pages}{339--367}.
\newblock


\bibitem[\protect\citeauthoryear{Box, Jenkins, Reinsel, and Ljung}{Box
  et~al\mbox{.}}{2015}]%
        {box2015time}
\bibfield{author}{\bibinfo{person}{George~EP Box}, \bibinfo{person}{Gwilym~M
  Jenkins}, \bibinfo{person}{Gregory~C Reinsel}, {and} \bibinfo{person}{Greta~M
  Ljung}.} \bibinfo{year}{2015}\natexlab{}.
\newblock \bibinfo{booktitle}{\emph{Time series analysis: forecasting and
  control}}.
\newblock \bibinfo{publisher}{John Wiley \& Sons}.
\newblock


\bibitem[\protect\citeauthoryear{Erf}{Erf}{2016}]%
        {Erf2016}
\bibfield{author}{\bibinfo{person}{Kyle Erf}.} \bibinfo{year}{2016}\natexlab{}.
\newblock \bibinfo{title}{Evergreen Continuous Integration: Why We Reinvented
  The Wheel}.
\newblock \bibinfo{howpublished}{Blog Post}.
\newblock
\urldef\tempurl%
\url{https://engineering.mongodb.com/post/evergreen-continuous-integration-why-we-reinvented-the-wheel}
\showURL{%
\tempurl}


\bibitem[\protect\citeauthoryear{Ingo and Daly}{Ingo and Daly}{2019}]%
        {Ingo2019}
\bibfield{author}{\bibinfo{person}{Henrik Ingo} {and} \bibinfo{person}{David
  Daly}.} \bibinfo{year}{2019}\natexlab{}.
\newblock \bibinfo{title}{Reducing variability in performance tests on EC2:
  Setup and Key Results}.
\newblock \bibinfo{howpublished}{Blog Post}.
\newblock
\urldef\tempurl%
\url{https://engineering.mongodb.com/post/reducing-variability-in-performance-tests-on-ec2-setup-and-key-results}
\showURL{%
\tempurl}


\bibitem[\protect\citeauthoryear{James and Matteson}{James and
  Matteson}{2015}]%
        {rpackage2007}
\bibfield{author}{\bibinfo{person}{Nicholas James} {and} \bibinfo{person}{David
  Matteson}.} \bibinfo{year}{2015}\natexlab{}.
\newblock \showarticletitle{ecp: An R Package for Nonparametric Multiple Change
  Point Analysis of Multivariate Data}.
\newblock \bibinfo{journal}{\emph{Journal of Statistical Software, Articles}}
  \bibinfo{volume}{62}, \bibinfo{number}{7} (\bibinfo{year}{2015}),
  \bibinfo{pages}{1--25}.
\newblock
\showISSN{1548-7660}
\urldef\tempurl%
\url{https://doi.org/10.18637/jss.v062.i07}
\showDOI{\tempurl}


\bibitem[\protect\citeauthoryear{Matteson and James}{Matteson and
  James}{2014}]%
        {edivisive2014}
\bibfield{author}{\bibinfo{person}{David~S. Matteson} {and}
  \bibinfo{person}{Nicholas~A. James}.} \bibinfo{year}{2014}\natexlab{}.
\newblock \showarticletitle{A Nonparametric Approach for Multiple Change Point
  Analysis of Multivariate Data}.
\newblock \bibinfo{journal}{\emph{J. Amer. Statist. Assoc.}}
  \bibinfo{volume}{109}, \bibinfo{number}{505} (\bibinfo{year}{2014}),
  \bibinfo{pages}{334--345}.
\newblock
\urldef\tempurl%
\url{https://doi.org/10.1080/01621459.2013.849605}
\showDOI{\tempurl}
\showeprint{https://doi.org/10.1080/01621459.2013.849605}


\bibitem[\protect\citeauthoryear{NIST/SEMATECH}{NIST/SEMATECH}{2012}]%
        {nisthandbook}
\bibfield{author}{\bibinfo{person}{NIST/SEMATECH}.}
  \bibinfo{year}{2012}\natexlab{}.
\newblock \bibinfo{title}{e-Handbook of Statistical Methods}.
\newblock
\newblock
\urldef\tempurl%
\url{http://www.itl.nist.gov/div898/handbook/}
\showURL{%
\tempurl}
\newblock
\shownote{note.}


\bibitem[\protect\citeauthoryear{Smart}{Smart}{2011}]%
        {smart2011jenkins}
\bibfield{author}{\bibinfo{person}{John~Ferguson Smart}.}
  \bibinfo{year}{2011}\natexlab{}.
\newblock \bibinfo{booktitle}{\emph{Jenkins: The Definitive Guide: Continuous
  Integration for the Masses}}.
\newblock \bibinfo{publisher}{" O'Reilly Media, Inc."}.
\newblock


\bibitem[\protect\citeauthoryear{Vallone}{Vallone}{2018}]%
        {GoogleWindowDetection}
\bibfield{author}{\bibinfo{person}{Anthony Vallone}.}
  \bibinfo{year}{2018}\natexlab{}.
\newblock \bibinfo{title}{Window Deviation Analyzer}.
\newblock
\newblock
\newblock
\shownote{Patent Publication 2018/0150373 A1, Filed Nov. 28, 2016, Published
  May 31, 2018.}


\bibitem[\protect\citeauthoryear{Walker-Morgan}{Walker-Morgan}{2019}]%
        {DJ2019}
\bibfield{author}{\bibinfo{person}{DJ Walker-Morgan}.}
  \bibinfo{year}{2019}\natexlab{}.
\newblock \bibinfo{title}{MongoDB 4.2 Performance Boosts}.
\newblock \bibinfo{howpublished}{Blog}.
\newblock
\urldef\tempurl%
\url{https://www.mongodb.com/blog/post/mongodb-42-performance-boosts}
\showURL{%
\tempurl}


\end{thebibliography}
\end{document}